\documentclass[12pt]{article} 
\usepackage[sectionbib]{natbib}
\usepackage{array,epsfig,fancyheadings,rotating}
\usepackage[]{hyperref}  
\usepackage{sectsty, secdot}
\sectionfont{\fontsize{12}{14pt plus.8pt minus .6pt}\selectfont}
\renewcommand{\theequation}{\thesection\arabic{equation}}
\subsectionfont{\fontsize{12}{14pt plus.8pt minus .6pt}\selectfont}

\usepackage{amsmath}
\usepackage{amssymb}
\usepackage{amsfonts}
\usepackage{multirow}
\usepackage{amsthm}

\theoremstyle{definition}

\usepackage{amsmath,amssymb,amsthm,amsfonts,bm,mathrsfs,float,algorithm,algorithmic,caption,subcaption}
\usepackage{avant,array,geometry,fontenc,inputenc,natbib,hyperref,multirow,enumerate,enumitem,rotating,booktabs,color,xcolor,latexsym,stmaryrd,xr}
\newtheorem{thm}{Theorem}

\newtheorem{pro}{Proposition}

\def\A{\boldsymbol{A}}

\def\H{\boldsymbol{H}}

\def\Q{\boldsymbol{Q}}
\def\R{\boldsymbol{R}}
\def\S{\boldsymbol{S}}

\def\Y{\boldsymbol{Y}}
\def\Z{\boldsymbol{Z}}

\def\a{\boldsymbol{a}}

\def\l{\boldsymbol{l}}
\def\s{\boldsymbol{s}}

\def\x{\boldsymbol{x}}

\def\0{\boldsymbol{0}}

\def\bbeta{\boldsymbol{\beta}}
\def\bfeta{\boldsymbol{\eta}}

\def\bdelta{\boldsymbol{\delta}}

\def\bR{\mathbb{R}}
\def\bS{\mathbb{S}}

\def\cI{\mathcal{I}}

\def\cP{\mathcal{P}}
\def\cS{\mathcal{S}}

\def\trans{^{\scriptscriptstyle \sf T}}

\newbox\TempBox \newbox\TempBoxA

\def\ep{\mathsf{E}} 
\def\id{\mathsf{I}} 
\def\tr{\mathsf{tr}} 
 
\def\Var{\mathsf{Var}} 
\def\Corr{\mathsf{Corr}}

\begin{document}


\renewcommand{\baselinestretch}{1.91}

\markright{ \hbox{\footnotesize\rm 
}\hfill\\[-13pt]
\hbox{\footnotesize\rm
}\hfill }

\markboth{\hfill{\footnotesize\rm FIRSTNAME1 LASTNAME1 AND FIRSTNAME2 LASTNAME2} \hfill}
{\hfill {\footnotesize\rm FILL IN A SHORT RUNNING TITLE} \hfill}

\renewcommand{\thefootnote}{}
$\ $\par


\fontsize{12}{14pt plus.8pt minus .6pt}\selectfont \vspace{0.8pc}
\centerline{\large\bf{NAPA: Neighborhood-Assisted and  Posterior-Adjusted }}
\vspace{2pt} 
\centerline{\large\bf Two-sample Inference}
\vspace{.4cm} 
\centerline{Li Ma$^\dag$, Yin Xia$^\dag$, and Lexin Li$^\ddag$} 
\vspace{.4cm} 
\centerline{\it $^\dag$Fudan University and $^\ddag$University of California at Berkeley}
 \vspace{.55cm} \fontsize{9}{11.5pt plus.8pt minus.6pt}\selectfont


\begin{quotation}
\noindent {\it Abstract:}
Two-sample multiple testing problems of sparse spatial data are frequently arising in a variety of scientific applications. In this article, we develop a novel neighborhood-assisted and posterior-adjusted (NAPA) approach to incorporate both the spatial smoothness and sparsity type side information to improve the power of the test while controlling the false discovery of multiple testing. We translate the side information into a set of weights to adjust the $p$-values, where the spatial pattern is encoded by the ordering of the locations, and the sparsity structure is encoded by a set of auxiliary covariates. We establish the theoretical properties of the proposed test, including the guaranteed power improvement over some state-of-the-art alternative tests, and the asymptotic false discovery control. We demonstrate the efficacy of the test through intensive simulations and two neuroimaging applications.

\vspace{9pt}
\par
\end{quotation}\par

\def\thefigure{\arabic{figure}}
\def\thetable{\arabic{table}}

\renewcommand{\theequation}{\thesection.\arabic{equation}}

\fontsize{12}{14pt plus.8pt minus .6pt}\selectfont

\section{Introduction}
\label{sec:introduction}

Two-sample hypothesis testing of sparse spatial data is a fundamental problem in a wide variety of scientific applications. It manifests in numerous forms. One example is to compare the cerebral white matter tracts between multiple sclerosis (MS) patients and healthy controls \citep{Goldsmith2011}. The data records the fractional anisotropy measure along the right corticospinal tract, and takes the form of one-dimensional (1D) function. The scientific interest is to compare two sets of fractional anisotropy profiles and locate the tract regions that distinguish cases from controls.  Another example is to compare the brain grey matter cortical thickness between subjects diagnosed with attention deficit hyperactivity disorder (ADHD) and typically developing controls \citep{ADHD200}. The data records the volume of grey matter at different brain locations in a three-dimensional (3D) space. The scientific interest is to compare two sets of brain structural images and identify differentiating brain regions. In addition to these examples, similar problems arise in many other applications, for instance, astronomical surveys \citep{Czakon2009}, disease mapping \citep{Sun2000}, ecology \citep{MauricioBini2009}, and genomics \citep{SunWei2011}. 

All these examples can be formulated as a two-sample testing problem, where the data reside in some spatial domain. More specifically, let $\bS \subset \bR^b$ denote a $b$-dimensional spatial domain, where $b=1,2,3,\ldots$. Let $\cS \subset \bS$ denote a finite, regular lattice in $\bS$, and $\s \in \cS$ the coordinate of the location. For the 1D MS example, $b=1$ and $\s$ is a scalar, whereas for the 3D ADHD example, $b=3$ and $\s$ is a three-variate coordinate. Later in our theoretical analysis, we consider the infill-asymptotic framework \citep{stein1999} and assume $\cS \to \bS$. Suppose the data $Y_d(\s) \in \bR$ is observed at every location $\s \in \cS$ for two groups $d=1,2$, and write $\Y_d = \{Y_d(\s) : \s \in \cS\}$. Suppose $\Y_d$ follows a probability distribution $\cP_{\bbeta_d,\bfeta_d}$, where $\bbeta_d=\{ \beta_{d}(\s) : \s \in \cS \}$ denotes the parameters of interest, and $\bfeta_d$ collects all the nuisance parameters. Suppose we observe two sets of independent samples, $\{\Y_{i,d} \}_{i=1}^{n_d}$, where $n_d$ is the sample size for group $d$, $d=1,2$. Our goal is to carry out multiple hypothesis testing given the observed data,
\begin{equation} \label{eqn:hypothesis}
H_0(\s): \; \beta_{1}(\s) = \beta_{2}(\s) \quad \mbox{versus} \quad H_1(\s): \; \beta_{1}(\s) \neq \beta_{2}(\s), \quad \s \in \cS.
\end{equation}
{We call the location $\s$ a signal location or a non-null location if $H_1(\s)$ holds, and call it a null location otherwise.} We comment that \eqref{eqn:hypothesis} covers a range of testing problems. In this article, we mostly illustrate with the problem of comparing two multivariate means \citep[e.g.,][]{tony2014two, xu2016adaptive}, where $\bbeta_d$ represents the mean of $\Y_d$. Meanwhile, our proposal is equally 
applicable to the problems of comparing large correlation or covariance structures {\citep{li2012two,cai2013two,cai2016large, zheng2019test}}, detecting differential networks \citep{xia2015testing, ChenKang2015}, or identifying gene-environment interactions \citep{Caspi2006, xia2018two}. Next, we recognize that, since the data resides in a spatial domain, there exists naturally some form of spatial smoothness in the data. Additionally, there is likely sparsity in \eqref{eqn:hypothesis}, in that the alternative hypothesis holds at only a small subset of locations $\s$ in the entire $\cS$. Sparsity is a common phenomenon in scientific applications and is frequently encountered in multiple testing. The goal of this article is to effectively incorporate both spatial smoothness and sparsity information into the multiple testing problem \eqref{eqn:hypothesis}, while controlling the false discovery rate (FDR) and improving the power of the test. 

Both smoothness and sparsity can be viewed as some forms of side information, and there have been a large number of proposals to incorporate side information in multiple testing; see \citet{CaiSun2017} for a review and references therein. In this article, we focus on the strategy of $p$-value weighting, which has been widely used for FDR control and power enhancement \citep[among many others]{BenHoc97, Storey2002, Genovese2006, Roeder2009}. In particular, \cite{Hu2010} adopted the prior knowledge that the hypotheses belong to a known number of groups, and weighed the $p$-values for the hypotheses in each group by $\pi_g / (1-\pi_g)$, where $\pi_g$ is the non-null proportion for the group that needs to be estimated. \cite{Zhang2011} proposed to smooth and aggregate the $p$-values in a local neighborhood to accommodate the spatial information of the neighboring $p$-values. \citet{Liu2014} utilized the sparsity information in the mean vectors and developed an uncorrelated screening-based FDR control procedure. \cite{Li2019} weighed the $p$-values by a heterogeneous weight $1/\{1-\hat\pi(s)\}$, where $\hat\pi(s)$ is the estimated probability of a {hypothesis} being a non-null. \cite{Ignatiadis2016, LeiFithian2018,lei2021general} incorporated generic side information through secondary data or external covariates, and constructed the $p$-value-based thresholding procedures adaptively. \citet{GAP} proposed the grouping, adjusting, and pooling (GAP) method that exploits the sparsity information, where they adaptively constructed a set of auxiliary statistics, based on which they identified clusters of hypotheses, then weighed the $p$-values by some discrete group-wise weights. \citet{LAWS} proposed the locally adaptive weighting and screening (LAWS) method that utilizes the smoothness information, where they constructed a set of robust and structure-adaptive weights based on the estimated local sparsity levels, then weighed the $p$-values by these continuous weights. 

In this article, we propose a \textbf{n}eighborhood-\textbf{a}ssisted and \textbf{p}osterior-\textbf{a}djusted test (NAPA) for the two-sample multiple testing problem \eqref{eqn:hypothesis}, whereas we aim to control the FDR and improve the power by incorporating both spatial and sparsity information. Our key idea is to translate the spatial and sparsity information to construct a set of weights to adjust the $p$-values, where the spatial pattern is encoded by the ordering of the locations, and the sparsity structure is encoded by a set of auxiliary statistics. {Specifically, we first construct a set of test statistics $\{T(\s) : \s \in \cS\}$, which contain useful information about the true signal locations $\theta(\s) = \id \{\beta_{1}(\s)\neq \beta_{2}(\s)\}$, and $\id(\cdot)$ is the indicator function. We then construct a set of auxiliary statistics $\{U(\s) : \s \in \cS\}$, which take the form of $\beta_1(\s) + \kappa(\s) \beta_2(\s)$ for some weight function $\kappa(\s)$, and contain useful information about $\gamma(\s) = \id\left\{ \beta_{1}(\s)\neq 0 \ \mbox{or}\ \beta_{2}(\s)\neq 0 \right\}$. We note that $\gamma(\s)=0$ implies $\theta(\s)=0$. This implication means that, if both $\beta_1(\s)$ and $\beta_2(\s)$ are zero, then the null hypothesis must be true, and at least one of $\beta_1(\s)$ and $\beta_2(\s)$ must be nonzero for the alternative hypothesis to hold. In other words, the locations where $\gamma(\s)=1$ capture useful information about the locations of the true signals. Moreover, the location $\s$ itself contains useful information about the smoothness embedded in the data. Therefore, we propose the posterior non-null probability, ${\pi(\s, U(\s)) = \Pr\{\theta(\s)=1|U(\s)\}}$, and weigh the $p$-value by the weight, {$w(\s,U(\s)) = \pi(\s,U(\s)) / \{1 - \pi(\s,U(\s)) \}$}, which integrates both the spatial information encoded in $\s$ and the sparsity information encoded in $U(\s)$.} Recognizing that $\pi(\s, U(\s))$ is a continuous function of $\s$ and $U(\s)$, we propose to estimate this posterior weight by pooling the information from its neighbors using a smoothing kernel approach. The neighborhood is defined through both $\s$ and $U(\s)$, in that a close spatial location in $\s$ and a similar value of $U(\s)$ both indicate a similar likelihood for the hypothesis at $\s$ to be null or alternative. Finally, given the estimated posterior weights, we choose a proper threshold to adjust the multiplicity for FDR control, and further show that the new test enjoys some guaranteed power improvement. 

Our proposed NAPA test is built on the recent proposals of multiple testing utilizing the side information. Particularly, our test combines the ideas of the GAP method of \citet{GAP} and the LAWS method of \citet{LAWS}. On the other hand, it is far from a straightforward extension, and is substantially different from both GAP and LAWS, as well as a simple combination of the two. More specifically, GAP translates the sparsity information embedded in the auxiliary statistic into several discrete groups, then constructs a weight for each group. However, the number of groups is generally unknown, and searching among all possible groupings is computationally expensive. Besides, such a discretization may lead to potential power loss. By contrast, our method does not weigh the $p$-values by groups, but instead weighs in a continuous fashion, which is computationally more efficient and can further improve the power. This new weighting strategy, nevertheless, induces new challenges. A key result in GAP, such that the grouping using auxiliary statistic does not distort the null distribution of the $p$-values, is no longer sufficient in our setting. To employ the auxiliary sequence continuously and to ensure the statistical properties, we derive a new conditional normal approximation, a result that is not available in the literature. Next, LAWS focuses on the one-sample testing problem and assumes that $\theta(\s)$ is fully determined by the location $\s$, but ignores additional sparsity information. By contrast, we introduce an auxiliary variable $U(\s)$ that is constructed adaptively from the data, and explore the posterior binomial variable $\theta(\s) | U(\s)$. The new approach, nevertheless, leads to a more involved theoretical development. This is because we have to tackle the correlations among the spatial locations and the covariates, as well as the dependencies among the primary and auxiliary statistics, and we derive new and sophisticated technical tools to address those challenges. Finally, our proposed test is far from a simple combination of GAP and LAWS. In Section \ref{sec:simulations}, we carry out a simulation experiment and show that our new test is much more powerful than naively combining GAP and LAWS. In summary, we believe our proposal fills an important gap in two-sample inference that utilizes both spatial and sparsity information, and thus helps address a range of scientific questions in areas such as neuroimaging analyses that involve spatial data. Moreover, we develop new technical tools that are potentially useful for general inference problems of complex dependent data.  

The rest of the article is organized as follows. Our proposed weight $w(\s,U(\s))$ and the posterior probability $\pi(\s, U(\s))$ hinge on both the location $\s$ and the auxiliary statistic $U(\s)$. In Section \ref{sec:oracle}, we first study this weighting scheme in the oracle setting, where $\pi(\s, U(\s))$ is known, the $p$-value obtained from the test statistic $T(\s)$ is uniformly distributed, and $U(\s)$ is independent of $T(\s)$ under the null. We establish the guaranteed power gain under this setting. In Section \ref{sec:napa}, we estimate the weight using smoothing kernel, and develop a multiple testing procedure based on the weighted $p$-values. We further illustrate the construction of the test and auxiliary statistics $\{T(\s), U(\s)\}$ in a multivariate mean comparison problem. In Section \ref{sec:theory}, we show that our weight estimator is consistent, and the proposed test controls the FDR asymptotically. We use the mean comparison as an example and show that the $p$-value from $T(\s)$ is {asymptotically} uniform, and $T(\s)$ and $U(\s)$ are asymptotically independent under the null.  In Section \ref{sec:simulations}, we study the finite-sample performance, and in Section \ref{sec:realdata}, we illustrate with two neuroimaging data examples. We relegate all technical proofs and additional simulations to the Supplementary Appendix.

\section{Oracle Weighting and Power Gain}
\label{sec:oracle}

In this section, we first derive and motivate our proposed weight function. We then study the power gain under the oracle setting.

\subsection{Posterior-adjusted weighting}
\label{subsec:weight}

Define the local non-null probability $\pi(\s)$ at location $\s$ as $\pi(\s) = \Pr\{\theta(\s)=1\}$, the local posterior non-null probability $\pi(\s,U(\s))$ at $\s$, and the weight function as, 
\begin{equation} \label{eqn:pi}
\begin{split}
\pi(\s,U(\s)) & = \Pr\{ \theta(\s)=1 | U(\s) \} = \frac{\pi(\s) \, q_{1}(U(\s)|\s)}{\{ 1-\pi(\s) \} \, q_{0}(U(\s)|\s) + \pi(\s) \; q_{1}(U(\s)|\s)}, \\
w(\s,U(\s)) & = \frac{\pi(\s,U(\s))}{1 - \pi(\s,U(\s))}, 
\end{split}
\end{equation}
where $q_{0}(\cdot | \s)$ and $q_{1}(\cdot | \s)$ are the null $\theta(\s) = 0$ and non-null $\theta(\s) = 1$ density functions of $U(\s)$, respectively. We call $\pi(\s,U(\s))$ a posterior probability as it is conditioning on the auxiliary variable $U(\s)$. {We propose to weigh the $p$-value $p(\s)$ as,}
\begin{equation} \label{eqn:weighted-p-value}       
p_w(\s) = \frac{p(\s)}{w(\s,U(\s))} = \frac{1-\pi(\s,U(\s))}{\pi(\s,U(\s))}p(\s).
\end{equation}

\subsection{Theoretical improvement}
\label{subsec:ranking}

We next compare our testing method with LAWS of \cite{LAWS} under the oracle setting. The two methods mainly differ in the weight used for the $p$-value, i.e., we use $w(\s, U(\s))$ in \eqref{eqn:pi}, whereas LAWS uses $w(\s) = \pi(\s) / \{1 - \pi(\s) \}$. Under some mild conditions, we show our method that incorporates both spatial and sparsity information is guaranteed to improve the power over LAWS that only utilizes the spatial information. We do not analytically compare with GAP of \cite{GAP} here, because GAP uses a discrete weighting scheme and is not directly comparable. Nevertheless, we numerically compare with LAWS and GAP in Section \ref{subsec:fdr-power}.

We first formally define the evaluation criteria in terms of false discovery and power. Consider a sequence of weighted $p$-values $\{ p_w(\s): s \in \cS\}$ and a given threshold $t$. When there is no weighting, we set all weights equal to one. Let $\delta^w(\s, t) = \id \left\{ p_w(\s) \leq t \right\}$ denote the decision rule for the hypotheses in \eqref{eqn:hypothesis}, in that $\delta^w(\s, t) = 1$ if we reject the null, and $\delta^w(\s, t) = 0$ otherwise. Let $\bdelta^w_t = \{\delta^w(\s, t) : \s \in \cS\}$ be the collection of all decision rules for $\s \in \cS$ under the threshold $t$. We define the FDR and the marginal FDR (mFDR) of the test $\bdelta^w_t$ as,
\begin{align*}
\mathrm{FDR}(\bdelta^w_t) & = \ep\left(\frac{\sum_{\s \in \cS}\left[\{1-\theta(\s)\} \delta^w(\s, t)\right]}{\max \left\{\sum_{\s \in \cS} \delta^w(\s, t), 1\right\}}\right), \\
\mathrm{mFDR}(\bdelta^w_t) & = \frac{\ep\left(\sum_{\s \in \cS}\left[\{1-\theta(\s)\} \delta^w(\s, t)\right]\right)}{\ep\left\{\sum_{\s \in \cS} \delta^w(\s, t)\right\}}.
\end{align*}
\cite{LAWS} showed that $\mathrm{FDR}(\bdelta^w_t) = \mathrm{mFDR}(\bdelta^w_t) + o(1)$ under some mild conditions. Therefore, we can use the leading term $\mathrm{mFDR}(\bdelta^w_t)$ to approximate $\mathrm{FDR}(\bdelta^w_t)$ asymptotically. In addition, we define the power of the test as
\begin{align*}
{\Psi(\bdelta^w_t) = \ep\left[\sum_{\s \in \cS} \left\{\theta(\s) \delta^w(\s, t)\right\}\right].}
\end{align*}
We further define the oracle threshold values of the two methods as, 
\begin{align*}
t_{\text{LAWS}} & = \sup_t \left\{ 0\leq t\leq 1: \textrm{mFDR}(\bdelta^{\text{LAWS}}_t) \leq \alpha \right\},\\
t_{\text{NAPA}} & = \sup_t \left\{ 0\leq t\leq 1: \textrm{mFDR}(\bdelta^{\text{NAPA}}_t) \leq \alpha \right\},
\end{align*}
where $\bdelta^{\text{LAWS}}_t$ and $\bdelta^{\text{NAPA}}_t$ represent the decision rules based on the LAWS and NAPA weights, respectively, and $\alpha$ is a pre-specified significance level.

Let $F_1$ denote the non-null conditional cumulative distribution function (CDF) of the {unweighted} $p$-value, and $F_1'$ its first derivative. Let $g_{\s}(x)=xF_1\{xt_{\text{LAWS}}/(1-x)|\s\}$ and $g'_{\s}(x)$ its first derivative. Let  $\pi_1=\text{Card}\left(\{\s\in\cS:\pi(\s)>0.5\}\right)/\text{Card}(\cS)$, where $\text{Card}(\cdot)$ is the cardinality of a set. The next theorem characterizes the theoretical gain of our NAPA method compared to LAWS.
{
\begin{thm} \label{thm1}
    For each $\s \in \cS$, suppose $U(\s)$ and $T(\s)$ are independent under the alternative. 
    Suppose $F_1(y|\s)$ is concave in $y$, $g_{\s}(x)$ is convex in $x$ for $x\leq 1/(1+t_{\text{LAWS}})$, and $yF_1'\left(y|\s\right)$ is non-decreasing in $y$. Suppose $\pi(\s)\in[\zeta,1-\zeta]$ for some small constant $\zeta>0$. If there exists some constant $0 < \varrho \leq 1$, such that $\pi_1 \leq \varrho$ and {$\{g'_{\s}(1-\zeta)-g'_{\s}(0.5)\}/\{1-g'_{\s}(0.5)\}\leq 1/\varrho$}, then
\begin{align*}
\begin{split}
    & \mathrm{mFDR}\left( \bdelta^{\text{NAPA}}_{t_{\text{LAWS}}} \right) \leq \mathrm{mFDR}\left( \bdelta^{\text{LAWS}}_{t_{\text{LAWS}}} \right) \leq \alpha, \\
    & \Psi\left( \bdelta^{\text{NAPA}}_{t_{\text{NAPA}}} \right) \geq \Psi\left( \bdelta^{\text{NAPA}}_{t_{\text{LAWS}}} \right) \geq \Psi\left( \bdelta^{\text{LAWS}}_{t_{\text{LAWS}}} \right).
\end{split}
\end{align*}
\end{thm}}
	
\noindent
We make a few remarks. First, Theorem \ref{thm1} shows that, when using the oracle threshold $t_{\text{LAWS}}$, our NAPA method achieves an mFDR that is no greater than that of LAWS and a power that is no smaller than that of LAWS. Based on this result and the construction of the oracle threshold, $t_{\text{NAPA}}$ is thus no smaller than $t_{\text{LAWS}}$. Therefore, using the threshold $t_{\text{NAPA}}$ leads to an additional power gain of NAPA, and thus it establishes the guaranteed power gain of NAPA over LAWS. Second, \cite{LAWS} showed that LAWS dominates the classical Benjamini and Hochberg (BH) method \citep{Benjamini1995}, and therefore, NAPA dominates BH too. {Third, this theorem does \emph{not} require any spatial dependence condition, as the power of the test is mainly reflected through the ranking of weighted $p$-values and the testing errors are evaluated by marginal FDR under the oracle setting when the weights are known. However, spatial dependence plays a crucial role in subsequent FDP and FDR control, as we show later in Section \ref{sec:theory}. Finally, the conditions in Theorem \ref{thm1} are all reasonably mild. In particular, as we show later when constructing the test statistic $T(\s)$ and the auxiliary variable $U(\s)$, they are asymptotically independent under both the null and alternative. The four distributional conditions can be easily verified for commonly used $p$-value distributions  \citep{sellke2001calibration, held2018p, zhang2022covariate}. Among them, similar concavity and convexity conditions have been commonly imposed in the FDR literature \citep[e.g.,][]{Storey2002, Genovese2006, Hu2010, GAP, LAWS}.  The condition $\pi_1\leq \varrho$ aligns with the sparsity framework we consider. We also give more discussion of the conditions in Theorem \ref{thm1} in Section \ref{subsec:verify-thm1} of the Appendix. }

\subsection{An illustration}
\label{subsec:illustration}

{Next, we consider a covariate-adjusted mixture model to offer an intuitive explanation of both our proposed weight, and the comparison with the case without using $U(\s)$. We clarify that we do \emph{not} impose such a model in our test, but only use it for illustration. Similar model has been frequently studied in the two-sample inference literature \citep{Efron2001, Newton2004, Sun2007, Efron2008}. 
\begin{align*}
{T(\s)|U(\s) \ {\sim} \ f(t|\s,U(\s))=\{1-\pi(\s,U(\s))\}f_0(t|\s,U(\s))+\pi(\s,U(\s))f_1(t|\s,U(\s)),}
\end{align*}
where $f_0$ and $f_1$ are the null and non-null conditional density function of $T(\s)$ given $U(\s)$, respectively. For the oracle setting, we have $f_0(t|\s,U(\s)) = f_0(t|\s)$. Following \cite{CARS}, when the tests are independent, the optimal test threshold for the above model is based on the ranking of the {conditional} local false discovery rate,
\begin{align*}
\mathrm{CLfdr}(t|\s,U(\s)) = \Pr\{\theta(\s) = 0 | t, \s, U(\s)\} \propto \frac{1-\pi(\s,U(\s))}{\pi(\s,U(\s))} \times \frac{f_0(t|\s)}{f_1(t|\s,U(\s))}.
\end{align*}
If we ignore $U(\s)$, the CLfdr reduces to the local false discovery rate, 
\begin{align*}
\mathrm{Lfdr}(t|\s) = \Pr\{\theta(\s) = 0 | t, \s\} \propto \frac{1-\pi(\s)}{\pi(\s)} \times \frac{f_0(t|\s)}{f_1(t|\s)}.
\end{align*}
When the tests are dependent, we consider weighing the $p$-value to approximate CLfdr or Lfdr. We observe that, in both CLfdr and Lfdr, the first term reflects the sparsity structure, whereas the second term reflects the strength of evidence against the null. However, the second term is usually difficult to estimate, so we replace it with the $p$-value. This essentially leads to our proposed weight $w(\s, U(\s))$ in \eqref{eqn:pi}, and the weight $w(\s) = \pi(\s) / \{1 - \pi(\s) \}$ used in LAWS. 

We also observe that, when comparing the two weights, a large $|U(\s)|$ usually provides a strong evidence that $\pi(\s, U(\s))$ is larger than $\pi(\s)$. This can be seen through the calculation, $\pi(\s,U(\s)) = \{\pi(\s) \, q_{1}(U(\s)|\s)\} / \big[ \{ 1-\pi(\s) \}$ $q_{0}(U(\s)|\s) + \pi(\s) \; q_{1}(U(\s)|\s) \big]$, where $q_{0}(\cdot | \s)$ and $q_{1}(\cdot | \s)$ are the null and non-null density functions of $U(\s)$. Therefore, by incorporating $U(\s)$, we may obtain a smaller weighted $p$-value for the alternatives and a better ranking of the tests, compared to LAWS that does not utilize $U(\s)$. Theorem \ref{thm1} then formally justifies such an intuition.}

\section{Two-sample Testing Procedure}
\label{sec:napa}

In this section, we first discuss how to estimate the posterior non-null probability {$\pi(\s, U(\s))$.} We then develop a general multiple testing procedure based on the weighted $p$-values. Finally, we illustrate the testing procedure with the problem of comparing two multivariate means, with a concrete construction of the test and auxiliary statistics $\{T(\s), U(\s)\}$.

\subsection{Neighborhood-assisted weight estimation}
\label{sec2}

Recognizing that it is rather difficult to directly estimate the posterior non-null probability $\pi(\s,U(\s))$ in \eqref{eqn:pi}, we first propose an intermediate quantity $\pi_{\tau}(\s,U(\s))$. We show that $\pi_{\tau}(\s,U(\s))$ provides a good approximation of $\pi(\s,U(\s))$, and the weight constructed based on $\pi_{\tau}(\s,U(\s))$ has the desired theoretical guarantees. A similar approximation has also been used in \cite{Schweder1982, Storey2002, LAWS}. Specifically, define  
\begin{equation} \label{eqn:intermediate}
\pi_{\tau}(\s,U(\s)) = 1-\frac{\Pr\big\{ p(\s)>\tau|U(\s)\big\}}{1-\tau}, \quad \textrm{ for some } \;\; 0<\tau<1.
\end{equation}
To justify the use of $\pi_{\tau}(\s,U(\s))$, consider 
\begin{align*}
\Pr\{p(\s) \leq t | U(\s)\} & = \{1-\pi(\s,U(\s))\} F_0(t|\s,U(\s))+\pi(\s,U(\s)) F_1(t | \s,U(\s)) \\
& \approx  \{1-\pi(\s,U(\s))\}t+\pi(\s,U(\s)) F_1(t |\s,U(\s)), 
\end{align*}
where $t\in[0,1]$, $F_0, F_1$ are respectively the {null and non-null conditional CDFs} of the $p$-value, and the approximation comes from the fact that the null $p$-value is asymptotically uniform and it is asymptotically independent of the auxiliary statistic $U(\s)$. Then the difference between $\pi_{\tau}(\s,U(\s))$ and $\pi(\s,U(\s))$ can be approximated by
$\{\pi_{\tau}(\s,U(\s))-\pi(\s,U(\s))\} / \pi(\s,U(\s)) \approx -\{1-F_1(\tau|\s,U(\s))\} / (1-\tau)$. Therefore, the difference between $\pi_{\tau}(\s,U(\s))$ and $\pi(\s,U(\s))$ is small with a properly chosen $\tau$, and it is asymptotically negative, which in turn would yield an asymptotically conservative FDR control. We discuss the choice of $\tau$ in Section \ref{subsec:estimation}. 

Next, we develop a neighborhood-assisted approach to estimate $\pi_\tau(\s,U(\s))$. Intuitively, the estimator can be obtained by counting the proportion of $p$-values that are greater than $\tau$ among all $p$-values at the location $\s$ and with the same auxiliary covariate value $U(\s)$. However, there is only one $p$-value at each $(\s,U(\s))$ pair. This prompts us to use a smoothing kernel approach to borrow information from the neighborhood of $(\s,U(\s))$. Specifically, consider a positive, bounded, unimodal kernel function $K(\x, y) : \bR^{b+1} \to \bR$ that is symmetric about zero in each dimension. Let $\H \in \bR^{(b+1)\times(b+1)}$ be a positive definite bandwidth matrix, and write {$K_{\H}(\s, U(\s)) = |\H|^{-1/2}$ $K\{\H^{-1/2}(\s\trans, U(\s))\trans\}$,} where $|\cdot|$ is the determinant. We briefly remark that, the bandwidth matrix $\H$ is not diagonal, because of the dependency between $\s$ and $U(\s)$ and possible correlations among the entries of $\s$.  If we set $\H$ as a diagonal matrix with the same magnitude along the diagonal and ignore $U(\s)$, then our estimator reduces to that in \cite{LAWS}. 
For a given $(\s,U(\s))$, we assign the weight to the ``pseudo-observation" of the $p$-value at $(\s',U(\s'))$ as
\vspace{-0.01in}
\begin{equation*}  
v_{\H}\{(\s,U(\s)),(\s',U(\s'))\}=\frac{K_{\H}(\s-\s',U(\s)-U(\s'))}{K_{\H}(\0,0)}.
\end{equation*}
Then, the number of ``pseudo-observations" that are greater than $\tau$ at each $(\s,U(\s))$ can be approximated by $\sum\nolimits_{\s'\in\cI(\tau)}v_{\H}\{(\s,U(\s)),(\s',U(\s'))\}$, where $\cI(\tau) = \{\s'\in\cS:p(\s')>\tau\}$. Meanwhile, the expectation of the $p$-values greater than $\tau$ can be calculated by $\left[\sum\nolimits_{\s'\in\cS} v_{\H}\{(\s,U(\s)),(\s',U(\s'))\}\right]$ $\left\{1-\pi_\tau(\s,U(\s))\right\}(1-\tau)$. Setting the two equal, we obtain an estimator of $\pi_\tau(\s,U(\s))$ as 
\begin{equation} \label{eqn:pi-tau}
\hat{\pi}_\tau(\s,U(\s)) = 1 - \frac{\sum_{\s'\in \cI(\tau)}v_{\H}\{(\s,U(\s)),(\s',U(\s'))\}}{(1-\tau)\sum_{\s' \in \cS}v_{\H}\{(\s,U(\s)),(\s',U(\s'))\}}.
\end{equation}
We obtain the neighborhood-assisted and posterior-adjusted weight estimator as 
\begin{equation} \label{eqn:weights}
\hat w(\s,U(\s)) = \frac{\hat{\pi}_\tau(\s,U(\s))}{1-\hat{\pi}_\tau(\s,U(\s))}.
\end{equation}

\subsection{Multiple testing procedure}
\label{subsec:testproc}

We next develop a general multiple testing procedure. We first observe that the expected number of false rejections with a known $\pi(\s, U(\s))$ at a given threshold $t$ can be computed as 
\begin{align*}
& {\ep_{\{p(\s),\theta(\s),U(\s)\},\s\in\cS}}\left[\sum_{\s\in\cS}\id \left\{p_w(\s)<t,\theta(\s)=0\right\}\right] \\
= & \sum_{\s\in\cS}\ep_{U(\s)}\left(\ep_{p(\s),\theta(\s)}\left[\id \left\{p_w(\s)<t,\theta(\s)=0\right\}|U(\s)\right]\right) \\
= & {\ep_{U(\s),\s\in\cS}}\sum_{\s\in\cS}\left[\Pr\{\theta(\s)=0|U(\s)\}\Pr\{p_w(\s)\leq t|\theta(\s)=0,U(\s)\}\right] \\
= & {\ep_{U(\s),\s\in\cS}} \sum_{\s\in\cS}\left[\{1-\pi(\s,U(\s))\}w(\s,U(\s))t\right]={\ep_{U(\s),\s\in\cS}}\left\{\sum_{\s\in\cS}\pi(\s,U(\s))t\right\}, 
\end{align*}   
in the oracle setting. If the $p$-value is uniformly distributed asymptotically, and $U(\s)$ is asymptotically independent of $T(\s)$ under the null, then for a given estimate $\hat{\pi}_\tau(\s,U(\s))$ and the decision rule $\id\{p_{\hat{w}}(\s)\leq t\}$, we can approximate the number of false rejection by $\sum_{\s\in\cS}\hat{\pi}_\tau(\s,U(\s))t$. We aim to reject as many hypotheses as possible, while controlling the estimated false discovery proportion (FDP) not to exceed the pre-specified significance level. This leads to the proposed testing procedure as summarized in Algorithm \ref{alg1}.

\begin{algorithm}[t!]
\caption{The multiple testing procedure of NAPA.}
\label{alg1}{
\begin{description}
\item[\textnormal{Step 1.}] Calculate the weights $\hat w(\s,U(\s))$ as in \eqref{eqn:weights}, and then adjust $p$-values by {$p_{\hat{w}}(\s)=p(\s)/\hat w(\s,U(\s))$} for $ \s \in \cS $.
			
\item[\textnormal{Step 2.}] Obtain the data-driven threshold 
\begin{align*}
{t}_{\hat{w}}=\sup_{t}\left\{0\leq t \leq 1: \frac{\sum_{\s \in \cS  } \hat{\pi}_{\tau}(\s,U(\s)) t}{\max \left\{\sum_{\s \in \cS  } \id\left\{p_{\hat{w}}(\s) \leq t\right\}, 1\right\}} \leq \alpha\right\}.
\end{align*}
			
\item[\textnormal{Step 3.}] Reject $H_{0}(\s)$ if $p_{\hat{w}}(\s) \leq {t}_{\hat{w}}$,  $\s\in\cS  $.
\end{description} }
\end{algorithm}

\subsection{{Comparison of multivariate means}}
\label{subsec:compare-means}

The problem of comparing two multivariate means has been widely studied, especially under the high-dimensional sparse setting {\citep[e.g.,][]{tony2014two, Liu2014, xu2016adaptive}.}
We illustrate the above general testing procedure with such a mean testing problem, and give a concrete construction of the test statistic $T(\s)$ and the auxiliary statistic $U(\s)$. Our method also applies to numerous other testing problems as well. 

{Specifically, given the observed data $\{Y_{i,d}(\s): \s \in \cS \}_{i=1}^{n_d}$, where $Y_{i,d}(\s)\sim \cP_{\beta_d(\s),\eta_d(\s)}$, $\beta_d(\s) =\ep\{Y_{i,d}(\s)\}$, $d=1,2$, we aim to test the hypotheses,}
\begin{equation*} 
{H_0(\s): \; \beta_{1}(\s) = \beta_{2}(\s) \quad \mbox{versus} \quad H_1(\s): \; \beta_{1}(\s) \neq \beta_{2}(\s), \quad \s \in \cS.}
\end{equation*} 
We construct the primary test statistic as
\begin{equation*}
T(\s) = \frac{\bar{Y}_1(\s)-\bar{Y}_2(\s)}{\left( \hat{\sigma}_{\s,1}^2 / n_1 + \hat{\sigma}_{\s,2}^2 / n_2 \right)^{1/2}}, \quad \s \in \cS, 
\end{equation*}
where $\bar{Y}_d(\s) = n_d^{-1} \sum_{i=1}^{n_d} Y_{i,d}(\s)$ is the group sample mean, and $\hat{\sigma}_{\s,d}^2 = n_d^{-1}\sum_{i=1}^{n_d}\{ Y_{i,d}(\s)$ $-\bar{Y}_d(\s) \}^2$ is the sample variance, $d=1,2$. Next, we construct the auxiliary statistic in the form of $\beta_1(\s) + \kappa(\s) \beta_2(\s)$, and for our mean comparison problem, we consider,  
\begin{equation*}
U(\s) = \frac{\bar{Y}_1(\s)+\hat{\kappa}(\s)\bar{Y}_2(\s)}{\left\{\hat{\sigma}_{\s,1}^2/n_1+\hat{\kappa}^2(\s)\hat{\sigma}_{\s,2}^2/n_2\right\}^{1/2}}, \quad \s \in \cS,
\end{equation*}
where $\hat{\kappa}(\s)=(n_2\hat{\sigma}_{\s,1}^2)/(n_1\hat{\sigma}_{\s,2}^2)$.

\section{Theoretical Properties}
\label{sec:theory}

In this section, we establish the theoretical properties of the NAPA testing procedure. We first show the estimated posterior non-null probability $\hat{\pi}_\tau(\s, U(\s))$ is a consistent estimator of $\pi_\tau(\s, U(\s))$. We then establish the asymptotic error rate control of NAPA under some conditions. Finally, we illustrate with the mean comparison problem again, and show the required conditions of NAPA are satisfied. Specifically, under the null, the test statistic $T(\s)$ is asymptotically normally distributed, and thus the corresponding $p$-value is asymptotically uniformly distributed, and $T(\s)$ is independent of the auxiliary statistic $U(\s)$ asymptotically. Throughout our asymptotic analysis, we consider the infill-asymptotic framework \citep{stein1999} that $\cS \to \bS$.

\subsection{Estimation consistency of the posterior probability}
\label{subsec:consistency}

We begin with some notations. Let $\cS_0=\{\s \in \cS : \theta(\s)=0\}$ and $\cS_1=\{\s \in \cS : \theta(\s)=1\}$ denote the set of null locations and non-null locations, respectively, and let $\cS = \cS_0 \cup \cS_1$. Let $m = \text{Card}( \cS )$, $m_0 = \text{Card}( \cS_0 ), m_1 = \text{Card}( \cS_1 )$, and $n = n_1 + n_2$.  For two  sequences of real numbers $\{a_{n}\}$ and $\{b_{n}\}$, write $a_{n} = O(b_{n})$ if there exists a constant $C$ such that $|a_{n}| \leq C|b_{n}|$ for any sufficiently large $n$, write $a_{n} = o(b_{n})$ if $\lim_{n\rightarrow\infty}a_{n}/b_{n} = 0$, and write $a_{n}\asymp b_{n}$ if there exists  constants $C>c>0$ such that $c|b_{n}| \leq |a_{n}| \leq C|b_{n}|$ for any sufficiently large $n$. Let $\lambda_{i}(\cdot)$ and $\tr(\cdot)$ denote the $i$th eigenvalue and the trace of a matrix, respectively.

Next, we show that the estimator $\hat{\pi}_\tau(\s,U(\s))$ in (\ref{eqn:pi-tau}) converges to the truth $\pi_{\tau}(\s,U(\s))$ for all $\s \in \cS$ as $\cS \rightarrow \bS$. {Let $\A \in \bR^{(b+2)\times (b+2)}$ denote the Hessian matrix of $\Pr\{p(\s')>\tau|U(\s'),U(\s)\}$ with respect to $(\s{'^\trans},U(\s'),U(\s))\trans$.} Partition the bandwidth matrix into $\H=\left(\begin{smallmatrix} \H_S & \a \\ \a\trans & h_U^2 \end{smallmatrix}\right)$, where $\H_S \in \bR^{b\times b}$, $\a \in \bR^{b\times 1}$, and $h_U^2 \in \bR$. Let $\tilde{h}=h_U^{2}-\a\trans\H_{\S}^{-1}\a\neq 0$. We introduce the following regularity conditions.

\vspace{-0.1in}
\begin{enumerate}[label=(C\arabic*), series=C]
\item \label{A0}  
Suppose the kernel function $K(\x,y):\bR^{b+1}\to \bR$ satisfies
\begin{align*} 
\int_{\bR^{b}} \x\trans\x K(\x,0)d\x < \infty,\int_{\bR^{b}} g^2(\x)K\{\Q\left[\x\trans,g(\x)\right]\trans\}d\x <\infty,    
\end{align*}
for any orthogonal matrix $\Q$ and any function $g:\bR^{b}\to \bR$.
	
\item \label{A1} Let $\Lambda_{\s}=\Big\{U(\s)$: with probability $1-O(m^{-1})$, uniformly for all $\s'\in\cS$, $\Pr(p(\s')>\tau|U(\s'),U(\s))$ has continuous first and second partial derivatives at $(\s{'^\trans},U(\s')$, $U(\s))\trans$ and $\lambda_i(\A)=O(1)$ for $i=1,\cdots,{b+2} \Big\}$. Suppose $\Pr(\Lambda_{\s})\rightarrow 1$ uniformly for all $\s\in\cS$ as $\cS\to\bS$.
	
\item \label{A2} {Suppose, uniformly for all $\s\in\cS$,
\begin{align*}
& \Var_{U(\s'),\s'\in\cS}\left(\sum_{\s' \in \cS  }\left[K_{\H}(\s-\s',U(\s)-U(\s'))\right]|U(\s)\right)= \\
& \hspace{1in}  O\left(\sum_{\s'\in \cS}\Var_{U(\s')}\left[ K_{\H}(\s-\s',U(\s)-U(\s'))|U(\s)\right] \right), \\
& \Var_{\{p(\s'),U(\s')\},\s'\in\cS}\left(\sum_{\s' \in \cS  }\left[K_{\H}(\s-\s',U(\s)-U(\s')) \id\left\{p(\s')>\tau\right\}\right]|U(\s)\right)= \\
& \hspace{1in}  O\left(\sum_{\s'\in \cS}\Var_{p(\s'),U(\s')}\left[ K_{\H}(\s-\s',U(\s)-U(\s')) \id\{p(\s')>\tau\}|U(\s)\right] \right)
\end{align*} 
hold with probability tending to 1 as $\cS\to\bS$.}
	
\item \label{A3} Suppose $\H_{\S}$ is nonsingular, {$\tilde{h}\neq 0$}, and uniformly for all $\s\in\cS$,
\begin{align*}
1/m \sum_{\s' \in \cS  } \ep_{U(\s')}\left[(\s-\s',U(\s)-U(\s'))\trans K_{\H}(\s-\s',U(\s)-U(\s'))|U(\s)\right]\to \0
\end{align*}
holds with probability tending to 1 as $\cS\to\bS$. Furthermore, suppose that
\begin{align*}
\tilde{h}=O(\tr(\H)), \tr(\H)=o(1), m^{-1}|{\H}|^{-1/2} = o(\tilde{h}^{1/2}).
\end{align*}
\end{enumerate}
\vspace{-0.1in}

\noindent
We make a few remarks about these conditions. Condition \ref{A0} holds for commonly used multivariate kernels, e.g., the standard normal kernel, the uniform kernel, among others. Condition \ref{A1} regulates the first and second derivatives of the conditional CDF of the $p$-values, and is mild. {Condition \ref{A2} assumes that most of the auxiliary statistics $\{U(\s)\}$ and most of the pairs $\{p(\s),U(\s)\}$ are weakly correlated across $\s$. This condition holds for numerous spatial structures. We give more discussion of this condition in Section \ref{subsec:verify-c3} of the Appendix.}  Condition \ref{A3} generally states the symmetry of the kernel function and is mild too. It can also be verified numerically. Besides, the requirement $m^{-1}|{\H}|^{-1/2} = o(\tilde{h}^{1/2})$ reduces to Condition (A2) in Lemma 1 of \cite{Duong2005} when ignoring the additional covariate $U(\s)$.

\begin{thm}\label{pro2}
Suppose Conditions \ref{A0} to \ref{A3} hold. Then, uniformly for all $\s \in \cS$, 
\begin{align*}
\hat{\pi}_{\tau}\left(\s,U(\s)\right)\to \pi_{\tau}\left(\s,U(\s)\right) \; \textrm {in probability, \ as } \; \cS \rightarrow \bS.
\end{align*}
\end{thm}

\noindent
We remark that, LAWS \citep{LAWS} only considered the spatial information and applied a diagonal smoothing kernel with a homogenous bandwidth to estimate the weight. In comparison, to integrate the neighborhood information encoded by both $\s$ and $U(\s)$, we develop a more sophisticated kernel estimation procedure that allows both non-orthogonal kernel components and heterogeneous bandwidth magnitudes.

\subsection{Asymptotic error rate control of NAPA}
\label{subsec:asymp-napa}

Next, we show the NAPA procedure controls both the FDR and the FDP asymptotically, where we define the FDP of the test $\bdelta_t^w = \{\delta^w(\s, t) : \s \in \cS\}$ as, 
\begin{equation*}
\mathrm{FDP}\left(\bdelta_t^{w}\right)=\frac{\sum_{\s \in \cS}\left[\{1-\theta(\s)\} \delta^{w}(\s, t)\right]}{\max \left\{\sum_{\s \in \cS} \delta^{w}(\s, t), 1\right\}}.
\end{equation*}
We again begin with some regularity conditions. 

\vspace{-0.1in}
\begin{enumerate}[resume*=C]
\item \label{A4} Suppose that $n_1\asymp n_2$, $\log m=o(n^{1/8})$. {Let $\psi(Y_{i,d}(\s))$ denote the influence function of $\beta_d(\s)$ at $Y_{i,d}(\s)$. Let $Z_{k}(\s) = (n_2/n_1)\psi(Y_{i,1}(\s))$, for $k = i, i=1, \ldots, n_1$, and $Z_{k}(\s) = -\psi(Y_{i,2}(\s))$, for $k = n_1 + i, i=1,\ldots,n_2$.} Suppose that $\ep\left\{\exp\left(C_1|Z_{k}(\s)|/[\Var\{Z_{k}(\s)\}]^{1/2}\right)\right\} < \infty$ for some $C_1 > 0$, and that there exists some ${\mu(\s)=\left(1+o\{(\log m)^{-1}\}\right)\ep\{U(\s)\}}$, such that 
\begin{align*}
&\Pr_{H_0(\s)}\left\{ \left|T(\s)-\frac{\sum_{k=1}^{n_1+n_2}Z_{k}(\s)}{\Var\{\sum_{k=1}^{n_1+n_2}Z_{k}(\s)\}^{1/2}}\right|\geq b_m \right\} = O(m^{-C_2}),\\
&\Pr_{H_0(\s)}\left\{\left|\left[U(\s)-{{\mu}(\s)}\right]-\frac{\sum_{k=1}^{n_1}Z_{k}(\s)-\vartheta(\s)\sum_{k=n_1+1}^{n_1+n_2}Z_{k}(\s)}{\Var\{\sum_{k=1}^{n_1}Z_{k}(\s)-\vartheta(\s)\sum_{k=n_1+1}^{n_1+n_2}Z_{k}(\s)\}^{1/2}}\right|\geq b_m\right\} = O(m^{-C_2}),
\end{align*}
where $\vartheta(\s) = \left[ n_1\Var\left\{Z_1(\s)\right\} \right] / \left[ n_2\Var\left\{Z_n(\s)\right\} \right]$, for some constant $C_2 > 5$ and $b_m = o\{(\log m)^{-1/2}\}$. 
	
\item \label{A6} For $\Z_k=\left\{Z_{k}(\s):\s\in\cS\right\}$ as defined in \ref{A4}, let $\R_1=\Corr(\Z_k)=\left(r_{\s,\boldsymbol{l};1}\right)_{m \times m}$ for $1\leq k\leq n_1$, $\R_2=\Corr(\Z_k)=\left(r_{\s,\l;2}\right)_{m \times m}$ for $n_1+1< k\leq n_1+n_2$, and suppose that $\max\{\lambda_{1}(\R_1),\lambda_{1}(\R_2)\}=o(m)$. Furthermore, let  $\Gamma_{\s}(\gamma)=\left\{\l: \l\in\cS,\left|r_{\s, \l;d}\right| \geq(\log m)^{-2-\gamma}, \ d=1\ \mbox{or}\ 2\right\}$, and suppose that there exists some $\gamma>0$, such that $\max_{\s\in\cS_0}\text{Card}\left\{ \Gamma_{\s}(\gamma) \right\} \asymp 1$.
	
\item \label{A7} Suppose, with probability tending to $1$, uniformly for all $\s\in \cS$, $\pi_{\tau}(\s,U(\s)) \in[\xi, 1-\xi]$ for some sufficiently small constant $\xi>0$, and $\pi_{\tau}(\s,U(\s))$ has bounded first derivatives with respect to $U(\s)$. Furthermore, suppose that 
\vspace{-0.01in}
\begin{equation*}
\Var_{\theta(\s),\s\in\cS}\sum_{\s\in\cS } \left(\ep_{U(\s)}\left[w(\s,U(\s))\big|\{\theta(\s):\s\in\cS\}\right]\id\left\{\theta(\s)=0\right\}\right)={o\left(m^2\right)}.
\end{equation*}

\item \label{A8} Let $\cS_{\nu}=\left\{\s\in\cS:\frac{\beta_1(\s)-\beta_2(\s)}{\Var\{\sum_{k=1}^{n}Z_{k}(\s)\}^{1/2}/n_2}\geq (\log m)^{1/2+\nu}\right\}$ {for $\Z_k=\left\{Z_{k}(\s):\s\in\cS\right\}$ as defined in \ref{A4}.} Suppose $\text{Card}\left(\cS_{\nu}\right) \geq \{1 / (c_\pi^{1 / 2} \alpha )+\varepsilon \} (\log m)^{1 / 2}$ for some $\varepsilon, \nu>0$, and $c_\pi$ is the ratio of a circle's circumference to its diameter. 
\end{enumerate}
\vspace{-0.1in}

\noindent
Condition \ref{A4} assumes the asymptotic normality of $T(\s)$ and $U(\s)$ under the null, which is easily attainable, as we illustrate with the problem of mean comparison in Section \ref{subsec:asymp-mean}. It also implies the asymptotic independence between $T(\s)$ and $U(\s)$ under the null, as we show in Lemma \ref{A5} of the Appendix. {Condition \ref{A6} requires that not too many variables have strong correlations that exceed $(\log m)^{-2-\gamma}$. This condition holds for numerous spatial structures. We give more discussion of this condition in Section \ref{subsec:verify-c6} of the Appendix.}  Condition \ref{A7} requires $\pi_{\tau}(\s, U(\s))$ to vary smoothly with respect to $U(\s)$, and not to be exactly 0 or 1 to ensure theoretical stability. It also assumes the latent variables {$\theta(\s)$} is {not perfectly correlated}, which ensures that $m_0 \asymp m$ has the probability tending to 1. Condition \ref{A8} requires a few spatial locations to have the standardized signal magnitude exceeding $(\log m)^{1/2+\nu }$, which avoids an overly conservative FDR. In general, these conditions are mild, and similar conditions of \ref{A6} to \ref{A8} have been imposed in \cite{LAWS}. 

The next two theorems establish the asymptotical control of FDR and FDP at the nominal level, first for a known $\pi_{\tau}(\s,U(\s))$ in Theorem \ref{thm2},  then for the estimated $\hat{\pi}_{\tau}(\s,U(\s))$ in Theorem \ref{thm3}. 

\begin{thm}\label{thm2}
Suppose Conditions \ref{A4} to \ref{A8} hold. Then,  
\vspace{-0.01in}
\begin{align*}
\varlimsup_{\cS \rightarrow \bS} \operatorname{FDR}\left(\boldsymbol{\delta}^{\text{NAPA}}_{t_w}\right) \leq \alpha, \;\; \text { and } \;\; 
\lim _{\cS \rightarrow \bS} \Pr\left\{\operatorname{FDP}\left(\boldsymbol{\delta}^{\text{NAPA}}_{t_w}\right) \leq \alpha+\epsilon\right\} = 1, \text{ for any } \; \epsilon > 0,
\end{align*}
where $t_{w}=\sup _{t}\left\{0\leq t\leq 1: \frac{\sum_{\s \in \cS  } \pi_{\tau}(\s,U(\s)) t}{\max \left\{\sum_{\s \in \cS  } \id\left\{p_{w}(\s) \leq t\right\}, 1\right\}} \leq \alpha\right\}$, and $w(\s,U(\s)) = \frac{\pi_\tau(\s,U(\s))}{1-\pi_\tau(\s,U(\s))}$.
\end{thm}

\begin{thm}\label{thm3}
Suppose Conditions \ref{A0} to \ref{A8} hold. Then,  
\begin{align*}
\varlimsup_{\cS \rightarrow \bS} \operatorname{FDR}\left(\boldsymbol{\delta}^{\text{NAPA}}_{t_{\hat w}}\right) \leq \alpha, \;\; \text { and } \;\; 
\lim _{\cS \rightarrow \bS} \Pr\left\{\operatorname{FDP}\left(\boldsymbol{\delta}^{\text{NAPA}}_{t_{\hat w}}\right) \leq \alpha+\epsilon\right\} = 1, \text{ for any } \; \epsilon > 0. 
\end{align*}
\end{thm}

\subsection{Asymptotic properties of mean comparison}
\label{subsec:asymp-mean}

Finally, we revisit the example of comparing multivariate means in Section \ref{subsec:compare-means}, and show that the required asymptotic normality and independence both hold. For other testing problems such as comparing the networks and detecting interactions, similar properties can be established accordingly; see \citet{GAP}. We introduce two additional regularity conditions. {Recall that, in this setting, $\beta_d(\s) =\ep\{Y_{i,d}(\s)\}$. Let $\sigma_{\s,d}^2 =\Var\{Y_{i,d}(\s)\}$, and $\kappa(\s) = (n_2\sigma_{\s,1}^2)/(n_1\sigma_{\s,2}^2)$, for all $\s \in \cS$.}

\vspace{-0.1in}
\begin{enumerate}[resume*=C]
\item \label{C1} Suppose $\log m = o(n^{1/5})$, $n_1\asymp n_2$, and $\sigma_{\s,1}^2\asymp \sigma_{\s,2}^2$ for all $\s \in \cS$.
	
\item \label{C2} There exists {some constant $C_1>0$,} such that $\ep[\exp \{C_1|Y_{i,d}(\s)-\beta_d(\s)|/\sigma_{\s,d}\}] < \infty$, for $d=1,2$ and all $\s \in \cS$.     
\end{enumerate}
\vspace{-0.1in}

\noindent 
Condition \ref{C1} allows the total number of hypotheses to test $m$ to grow exponentially with the total sample size $n$, while requiring the sample size and the variance of each group to be of the same order. Condition \ref{C2} holds for a broad family of distributions with an exponential tail and is similarly assumed in various testing literatures \citep[e.g.,][]{cai2013two,cai2016large,guo2021specification,he2021asymptotically}. Both conditions are mild.

\begin{pro}\label{pro1}
Suppose Conditions \ref{C1} and \ref{C2} hold. Then under the null hypothesis, for any constant $C_3>0$, there exists ${b_m} = o\left\{(\log m)^{-1/2}\right\}$, such that 
\begin{eqnarray*}
&& \Pr\left\{\left|{T}(\s)-\frac{\bar{Y}_1(\s)-\bar{Y}_2(\s)}{\left(\sigma_{\s,1}^2/n_1+\sigma_{\s,2}^2/n_2\right)^{1/2}}\right|\geq b_m\right\} = O\left( m^{-C_3} \right), \\
&&\Pr\left\{\left|\left[U(\s)-{\mu(\s)}\right]-\frac{\bar{Y}_1(\s)-\beta_1(\s)+\kappa(\s)\{\bar{Y}_2(\s)-\beta_2(\s)\}}{\left\{\sigma_{\s,1}^2/n_1+\kappa^2(\s)\sigma_{\s,2}^2/n_2\right\}^{1/2}}\right|\geq b_m\right\} = O(m^{-C_3}),
\end{eqnarray*} 
uniformly for $\s\in\cS_0$, where ${\mu}(\s)=\left(1+o\{(\log m)^{-1}\}\right)\ep\{U(\s)\}$. This further implies that, for any constant $C_4>0$,
\begin{align*}
\Pr\left\{|T(\s)|\geq t|U(\s)\right\}=\{1+o(1)\}G(t)+O(m^{-C_4}),
\end{align*}
uniformly in $t = O\{(\log m)^{1/2}\}$, {$|U(\s)-{\mu(\s)}| = O \{(\log m)^{1/2}\}$, and all $\s \in \cS_0$,} where $G(t)=2\{1-\Phi(t)\}$, and $\Phi(\cdot)$ is the CDF of a standard normal random variable.
\end{pro}

\noindent
We make some remarks. First, \cite{GAP} illustrated their GAP test with the problem of comparing two multivariate means too, but they only studied the multivariate normal distribution, while Proposition \ref{pro1} extends to the family of distributions with an exponential tail. Second, compared to the asymptotic independence result in GAP, Proposition \ref{pro1} establishes the exact conditional probability tail of $T(\s)$ given $U(\s)$. As a result, the proof of Proposition \ref{pro1} is technically much more involved. Toward our goal, we obtain a  conditional normal approximation result in the proof of Proposition \ref{pro1}, which to our knowledge is not available in the literature. Finally, because we incorporate the auxiliary statistic in the construction of the weight, Proposition \ref{pro1} ensures that the null distribution of $T(\s)$ is not to be affected by the observed $U(\s)$. By contrast, the auxiliary statistic is only used for the grouping purpose in GAP, and hence their asymptotic independence result can be viewed as a simpler discretized version of Proposition \ref{pro1}.

\section{Simulations}
\label{sec:simulations}

In this section, we first study the capability of NAPA in recovering the posterior non-null probability $\pi(\s, U(\s))$, then the finite-sample performance of NAPA, and compare with BH \citep{Benjamini1995}, GAP \citep{GAP}, LAWS \citep{LAWS}, and a simple combination of GAP and LAWS. This combination method first applies GAP with three groups to obtain the group-wise reweighted $p$-values, then feeds into the LAWS method for the second-stage reweighing. We present additional simulations in Section \ref{subsec:more-simu}, and study irregular domain and lattice in Section \ref{subsec:irregular}, and heavy-tailed distribution in Section \ref{subsec:heavytail} of the Appendix.

\subsection{Posterior probability estimation}
\label{subsec:estimation}

Given the key role the posterior non-null probability plays, we first evaluate the capability of NAPA in recovering $\pi(\s, U(\s))$ through the estimator $\hat{\pi}_\tau(\s,U(\s))$ in \eqref{eqn:pi-tau}. We simulate two groups of independent samples $\{Y_{i,1}(\s)\}_{i=1}^{n_1}$ and $\{Y_{i,2}(\s)\}_{i=1}^{n_2}$, 
\vspace{-0.01in}
\begin{align} \label{eqn:sim}
\begin{split}
Y_{i,1}(\s) \ | \ \theta(\s) & \sim  \{ 1-\theta(\s) \} \ \text{Normal}(0,1) \ + \ \theta(\s) \ \text{Normal}\big( {\beta_1(\s)},1 \big),\\
Y_{i,2}(\s) \ | \ \theta(\s) & \sim  \{ 1-\theta(\s) \} \ \text{Normal}(0,4) \ + \ \theta(\s) \ \text{Normal}\big( {\beta_2(\s)},4 \big),       
\end{split}
\end{align}
where $\theta({\s}) \sim \text{Bernoulli}(1, \pi({\s}))$, $\beta_1(\s) = 1 / \sqrt{20}$, and {$\beta_2(\s) = 2/\sqrt{5}$}. Note that $\pi(\s)$ specifies the likelihood of possible signal locations. We consider three examples of generating the signal regions: a 1D example of a piecewise constant-shaped signal, a 2D example of two rectangular-shaped signals, and a 3D example of a cubic-shaped signal. For the 1D case, we consider $s = 1, 2, \ldots, 5000$, and we set $\pi(s)=0.8$ for $s \in [1001, 1200] \cup [2001, 2200]$, and $\pi(s)=0.6$ for $s \in [3001, 3200] \cup [4001, 4200]$. For the 2D case, we consider $\s = (s_1, s_2)$, with $s_1 = 1, 2, \ldots, 100, s_2 = 1, 2, \ldots, 50$, and we set $\pi(\s)=0.8$ for the left signal rectangle when $s_1 \in [20, 40], s_2 \in [10, 30]$, and $\pi(\s)=0.6$ for the right signal rectangle when $s_1 \in [60, 80], s_2 \in [10, 30]$. For the 3D case, we consider $\s = (s_1, s_2, s_3)$, with $s_1 = 1, 2, \ldots, 20, s_2 = 1, 2, \ldots, 25, s_3 = 1, 2, \ldots, 15$, and we set $\pi(\s)=0.7$ for the signal cube when $s_1 \in [5, 15], s_2 \in [5, 15], s_3 \in [1, 10]$. We set $\pi(\s)=0.05$ for all the rest of locations. We set the sample size at $n_1 = n_2 = 100$. Given the generative model \eqref{eqn:sim}, we can derive the explicit distribution of $U(\s) | \theta(\s)$, and plugging it into \eqref{eqn:pi} yields the true posterior non-null probability $\pi(\s, U(\s))$. 

\begin{figure}[t!]
\centering
\renewcommand{\baselinestretch}{1}
\includegraphics[width=0.95\linewidth, height=3.1in]{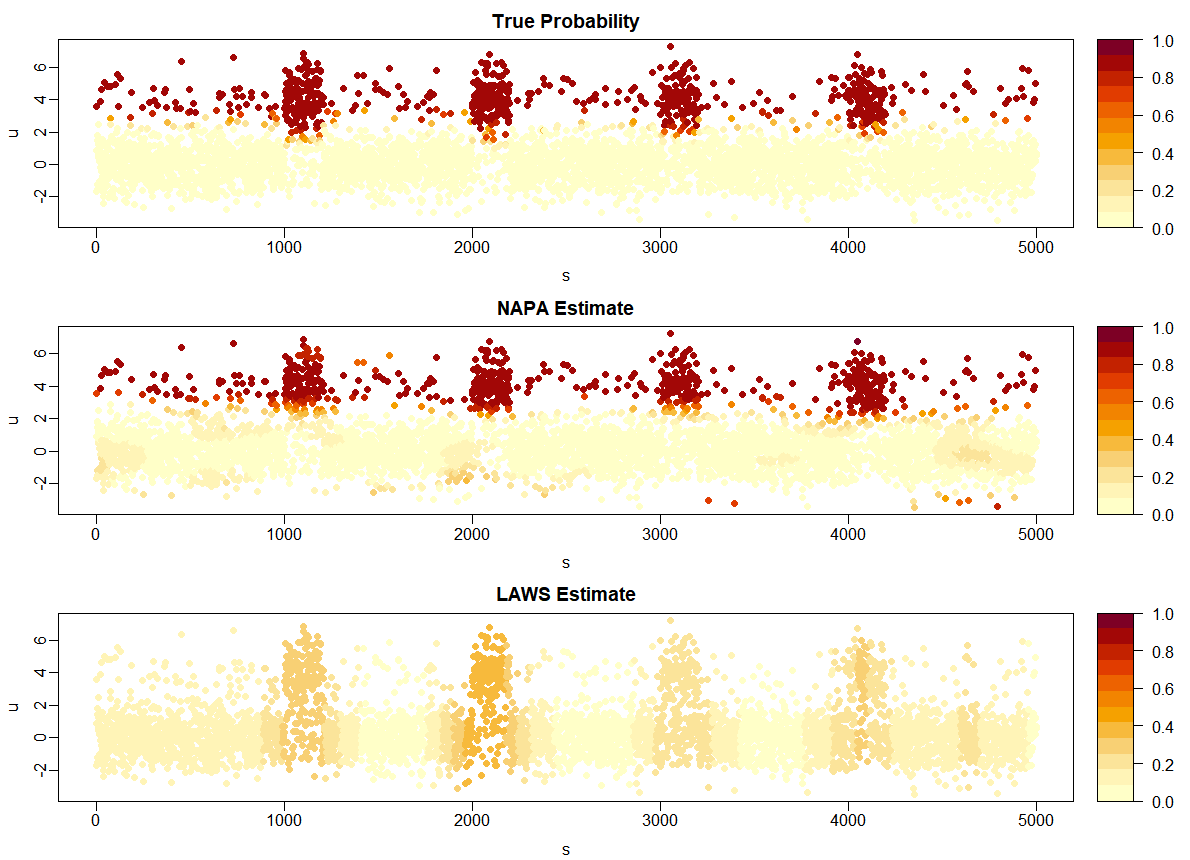}
\caption{\small Estimation of the posterior non-null probability $\pi(\s, U(\s))$ for the 1D example. From top to bottom: the true probability, the estimated probability by NAPA, and the estimated probability by LAWS.}   
\label{fig:1D-pi}
\end{figure}

Next, we estimate $\pi(\s,U(\s))$ using a bivariate Gaussian kernel function with the two-dimensional bandwidth matrix,
\begin{align} \label{eqn:bandwidthH}
\H=\begin{pmatrix}
h_{\s}^2 & \rho h_{\s}h_U\\
\rho h_{\s}h_U & h_U^2 
\end{pmatrix},
\end{align}
where $h_{\s}$ and $h_U$ are the bandwidths for $\| \s - \s' \|$ and $\| U(\s) - U(\s') \|$, respectively, $\rho$ is the correlation between $\| \s - \s' \|$ and $\| U(\s) - U(\s') \|$, and $\| \cdot \|$ denotes the Euclidean norm. {We use the plug-in selector \citep{sheather1991reliable} and the normal-scale selector \citep{chacon2011asymptotics} to obtain $h_{\s}$ and $h_U$, respectively.
Specifically, we employ the \texttt{R} package \texttt{ks} \citep{Duong2007, duong2022ks}, in which we use the function \texttt{hpi} to select the bandwidth for $\| \s - \s' \|$, and the function  \texttt{hns} to select the bandwidth for $\| U(\s) - U(\s') \|$. We estimate $\rho$ by the sample correlation. We further carry out a sensitivity analysis for the bandwidth selection in Section \ref{subsec:sensitivity} of the Appendix.} Moreover, we follow \cite{LAWS} and choose $\tau$ in \eqref{eqn:pi-tau} as the cutoff $p$-value when applying the BH procedure to the sequence of all the unweighted $p$-values at the significance level $0.9$. This ensures that the null cases are dominant in the set $\cI(\tau) = \{\s'\in\cS:p(\s')>\tau\}$. Finally, to stabilize the probability estimation, we truncate $\hat{\pi}_{\tau}(\s,U(\s)) = \xi$ if $\hat{\pi}_\tau(\s,U(\s))<\xi$, and $\hat{\pi}_{\tau}(\s,U(\s)) = 1-\xi$ if $\hat{\pi}_\tau(\s,U(\s))>1-\xi$, where we set $\xi=10^{-5}$.  

\begin{figure}[t!]
\centering
\renewcommand{\baselinestretch}{1}
\includegraphics[width=0.95\linewidth, height=1.2in]{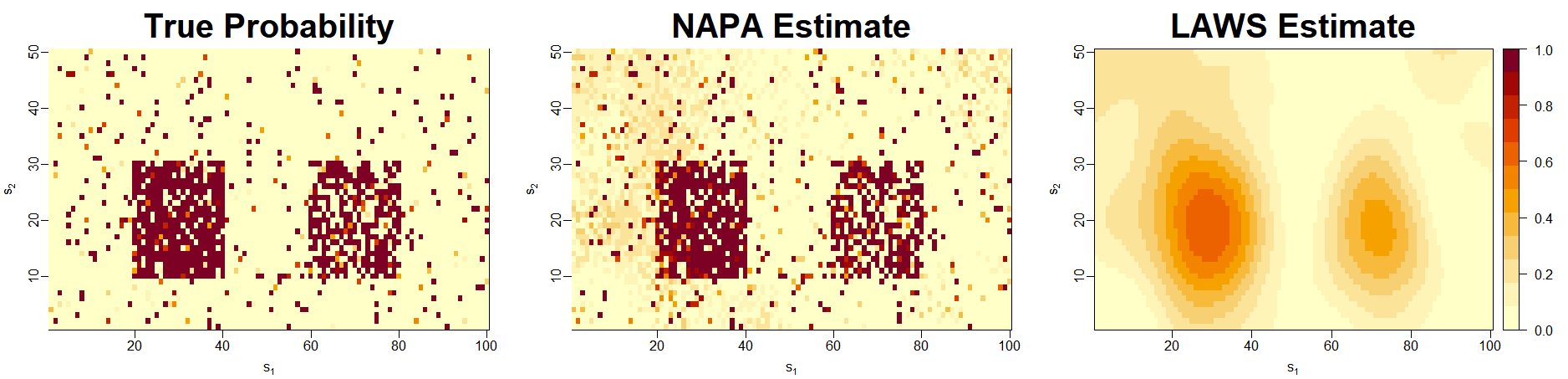}
\caption{\small Estimation of the posterior non-null probability $\pi(\s, U(\s))$ for the 2D example. From left to right: the true probability, the estimated probability by NAPA, and the estimated probability by LAWS.}
\label{fig:2D-pi}
\end{figure}

\begin{figure}[t!]
\centering
\renewcommand{\baselinestretch}{1}
\includegraphics[width=0.95\linewidth, height=2.85in]{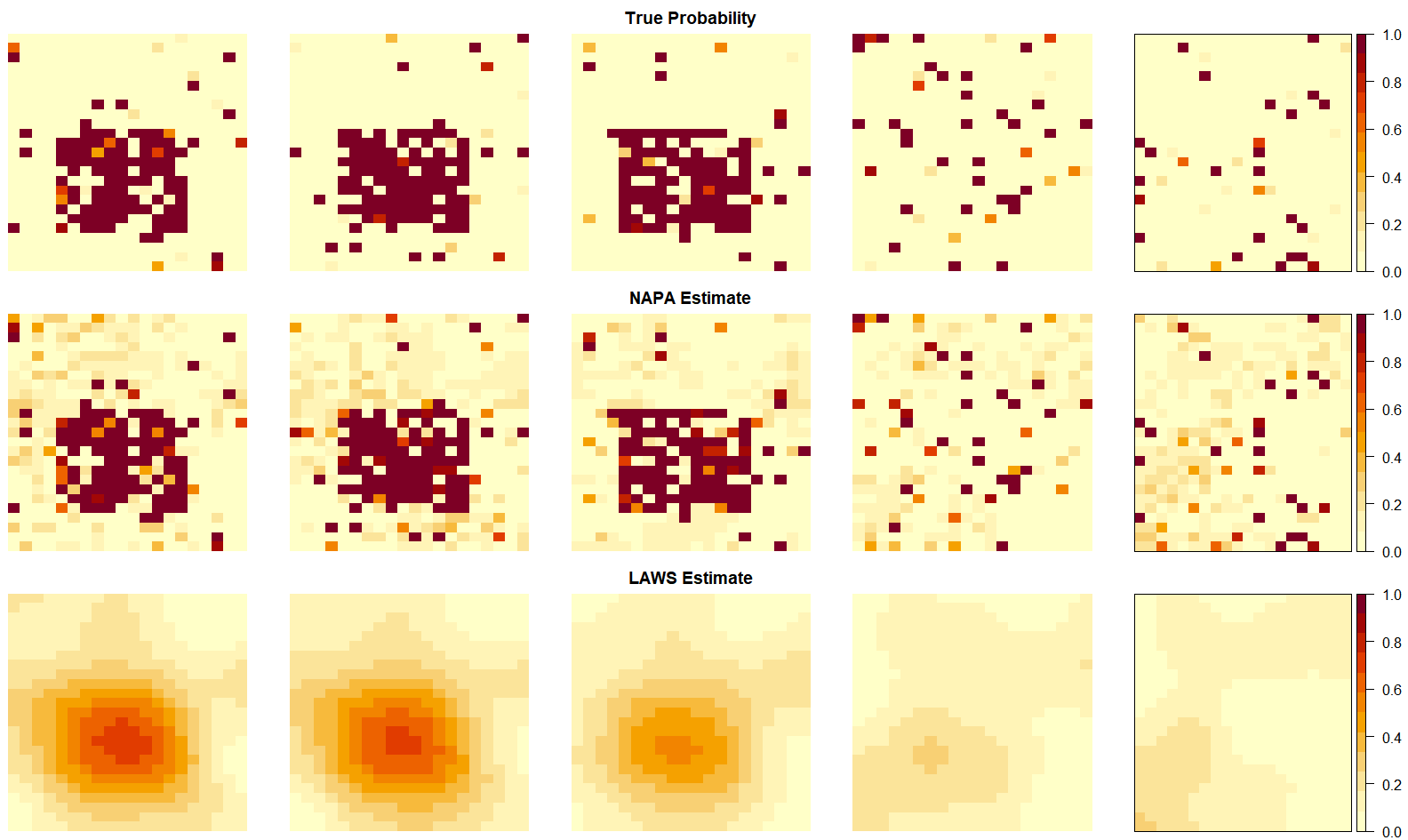}
\caption{\small Estimation of the posterior non-null probability $\pi(\s, U(\s))$ for the 3D example. From left to right: five selected slices with $s_3 = 3, 6, 9, 12, 15$. From top to bottom: the true probability, the estimated probability by NAPA, and the estimated probability by LAWS.}
\label{fig:3D-pi}
\end{figure}

We compare our posterior probability estimator $\hat{\pi}_\tau(\s,U(\s))$ in \eqref{eqn:pi-tau} that utilizes both the smoothness and sparsity information with the truth ${\pi}(\s,U(\s))$. We also compare with the corresponding estimator $\hat{\pi}_\tau(\s)$ used in LAWS that only utilizes the smoothness information alone. Figures \ref{fig:1D-pi}, \ref{fig:2D-pi} and \ref{fig:3D-pi} report the results based on a single data replication for the 1D, 2D and 3D examples, respectively. It is clearly seen that our posterior probability estimator $\hat{\pi}_\tau(\s,U(\s))$ is much closer to the truth than the LAWS estimator $\hat{\pi}_\tau(\s)$. 

\subsection{FDR and power comparison}
\label{subsec:fdr-power}

Next, we evaluate the empirical FDR and power of the proposed NAPA method, and compare it with BH, GAP, LAWS, and a simple combination of GAP and LAWS. 

\begin{figure}[t!]
\centering
\renewcommand{\baselinestretch}{1}
\includegraphics[width=0.95\linewidth, height=3.25in]{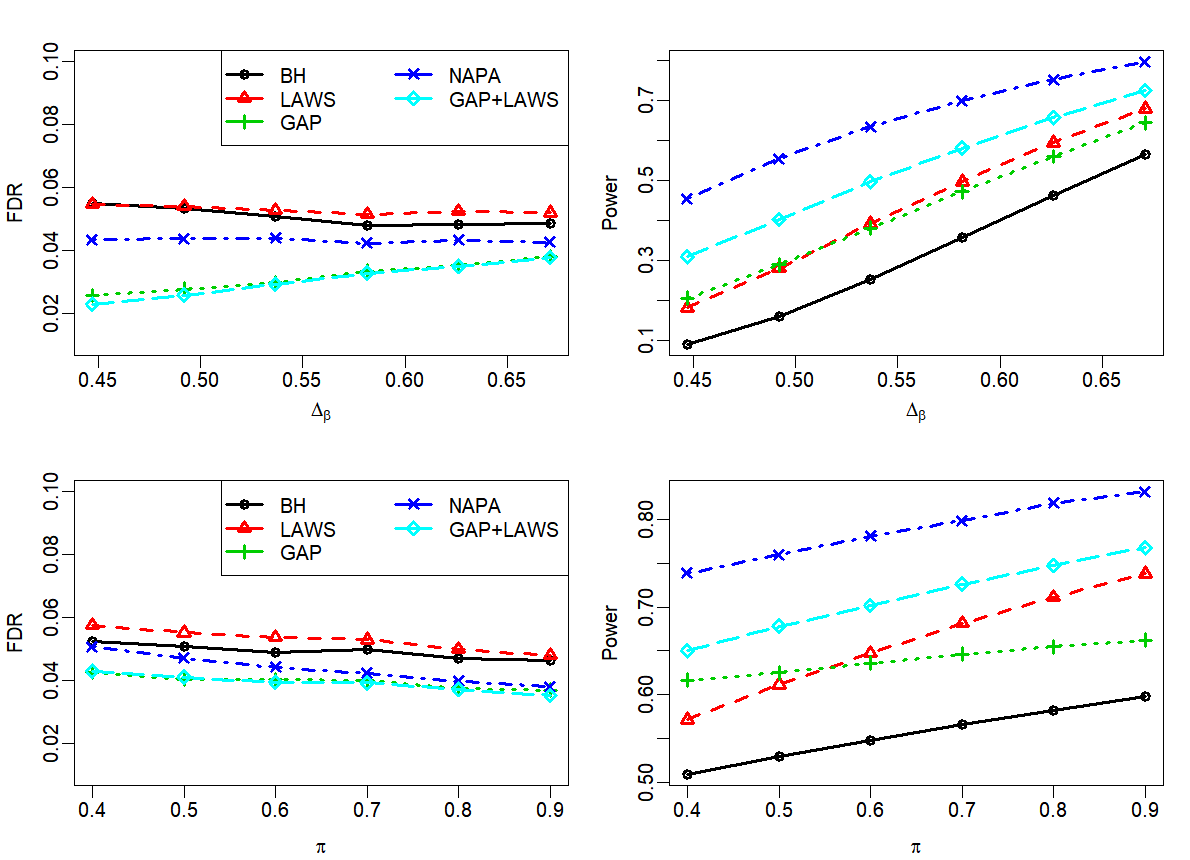}
\caption{\small Empirical FDR and power for the 1D example. Top panels: varying $\Delta_{\beta}$ in scenario 1, and bottom panels: varying $\pi$ in scenario 2. Five methods are compared: the proposed method (NAPA), the GAP method \citep{GAP}, the LAWS method \citep{LAWS}, the simple combination of GAP and LAWS, and the BH method \citep{Benjamini1995}.}
\label{fig:1D}
\end{figure}

We continue to simulate the data from model \eqref{eqn:sim} with the 1D, 2D and 3D examples. {Denote by $\Delta_{\beta}=\beta_2(\s)-\beta_1(\s)$ and it has the same value across all signal locations. Note that $\Delta_{\beta}$ controls the strength of the signal,} whereas $\pi(\s)$ specifies the likelihood of possible signal locations. We consider two scenarios: vary $\Delta_{\beta}$ from $1/\sqrt{5}$ to $3/\sqrt{20}$, while fixing $\pi(\s)$ in the same way as in Section \ref{subsec:estimation}; {vary $\pi(\s)=\pi$ in all signal regions from $0.4$ to $0.9$,} while fixing $\Delta_{\beta}=3/\sqrt{20}$. Let $\beta_1(\s) \sim \text{Uniform}(-1,1)/\sqrt{5}$ in both scenarios. We set the sample size at $n_1 = n_2 = 100$, and set the nominal level at $\alpha = 0.05$.

\begin{figure}[t!]
\centering
\renewcommand{\baselinestretch}{1}
\includegraphics[width=0.95\linewidth, height=3.25in]{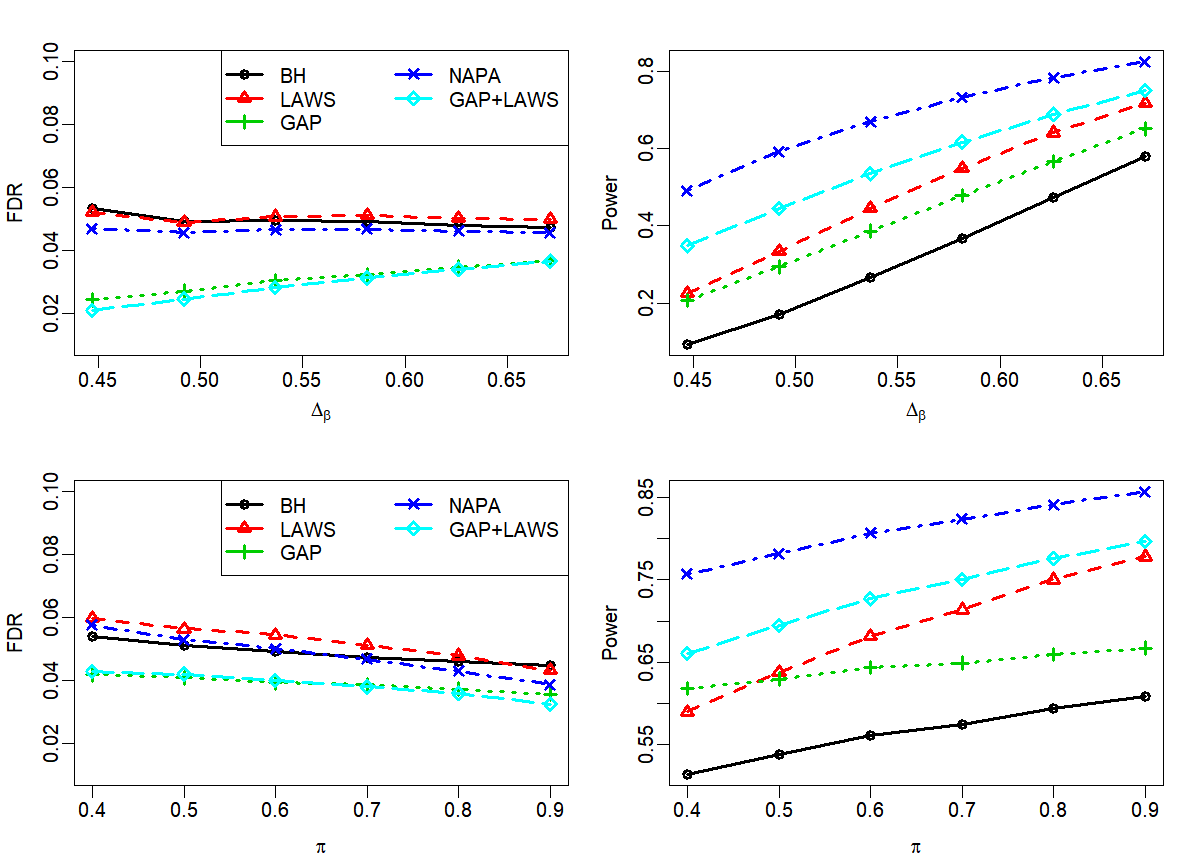}
\caption{\small Empirical FDR and power for the 2D example, with same legend as Figure \ref{fig:1D}.}
\label{fig:2D}
\end{figure}

\begin{figure}[t!]
\centering
\renewcommand{\baselinestretch}{1}
\includegraphics[width=0.95\linewidth, height=3.25in]{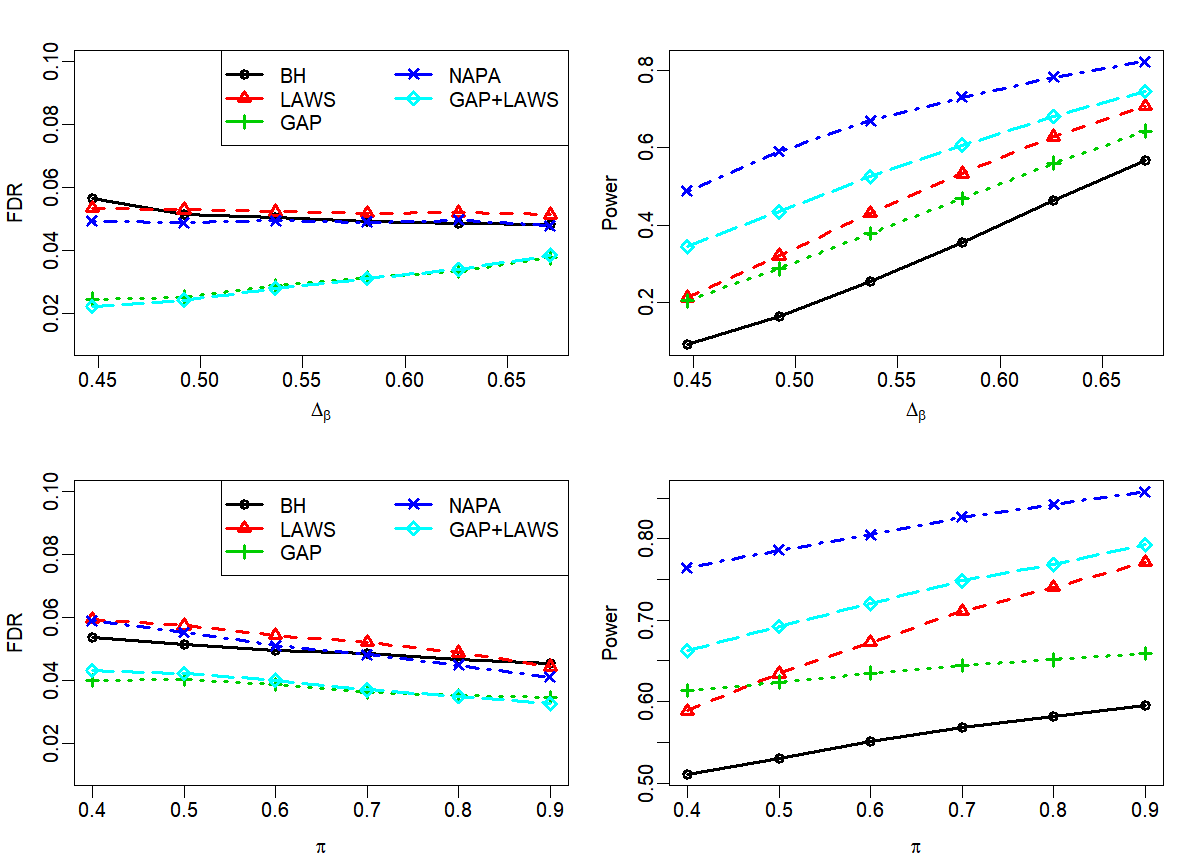}
\caption{\small Empirical FDR and power for the 3D example, with same legend as Figure \ref{fig:1D}.}
\label{fig:3D}
\end{figure}

Figures \ref{fig:1D}, \ref{fig:2D}, and \ref{fig:3D} report the empirical FDR and power of various testing methods based on 200 data replications. It is clearly seen that, in all three examples, while all methods can control the FDR {around the nominal level}, our proposed NAPA method achieves the most power gain compared to all the alternative methods. These results agree with our theory as well as our intuition that the NAPA method that utilizes both spatial and sparsity information outperforms the GAP and LAWS methods that utilize only one type of side information alone. Moreover, our method clearly outperforms the simple combination of GAP and LAWS. This simple combination has no theoretical guarantee. Besides, it loses information when applying the GAP method that reweighs the $p$-values in a discrete fashion. Through these examples, we see that our proposed NAPA test is more than just a simple combination of GAP and LAWS.

\section{Real Data Applications}	

\label{sec:realdata}

In this section, we illustrate our proposed test with two neuroimaging applications.

\subsection{Multiple sclerosis study}
\label{subsec:real-1D}

The first study is to compare the cerebral white matter tracts between multiple sclerosis (MS) patients and healthy controls \citep{Goldsmith2011}. MS is a demyelinating autoimmune disease that causes lesions in the white matter tracts of a patient and results in severe disability. Diffusion tensor imaging (DTI) is a magnetic resonance imaging (MRI) technique that studies white matter tractography by measuring the diffusivity of water in the brain. The data records the fractional anisotropy measure, which describes the degree of diffusion anisotropy, along the right corticospinal tract for $n_1 = 340$ multiple sclerosis patients and $n_2 = 42$ healthy controls. The tract data are generally modeled as 1D functions, and there are in total $|\cS| = 43$ locations for each tract. The dataset is available in the \texttt{R} library \texttt{refund}, and the data processing information can be found in \citet{Luo2017}. {The scientific interest here is to compare the two mean  functional profiles of diffusivity and locate the tract locations that distinguish cases from controls.}  

We apply the NAPA test to this dataset, and also compare it with the alternative tests, all under the nominal level $\alpha = 0.05$. The number of identified differential locations by NAPA, BH, GAP, LAWS, and their simple combination is {30, 0, 7, 25, and 17,} respectively. Besides, the set of locations found by NAPA is a superset of those found by GAP and the simple combination of GAP and LAWS, and is also a superset of those found by LAWS except for two locations. Together with our simulation studies, it seems to suggest that our proposed NAPA test manages to achieve the best power. It is also interesting to note that the locations identified by NAPA concentrate on the regions with distances 13 to 20, 22 to 32, and 37 to 47 along the tract. Such a finding warrants additional scientific validation.

\subsection{Attention deficit hyperactivity disorder study}
\label{subsec:real-3D}

The second study is to compare the brain grey matter cortical thickness between subjects diagnosed with attention deficit hyperactivity disorder (ADHD) and typically developing controls \citep{ADHD200}. ADHD is one of the most common child-onset neurodevelopmental disorders. Anatomical MRI is an imaging technique that studies brain anatomical structures. The data records the volume of grey matter at different brain locations in a 3D space for $n_1 = 356$ ADHD subjects and $n_2 = 575$ normal controls. The dataset is available at \url{http://neurobureau.projects.nitrc.org/ADHD200/Data.html}. The MRI images were preprocessed by the Neuro Bureau using the burner pipeline \citep{ADHD200}. To reduce the dimensionality of the problem, we further downsize the image resolution from $256 \times 198 \times 256$ to $30 \times 36 \times 30$, following the same data reduction strategy as in \citet{Li2017} and \citet{LAWS}. {The scientific interest here is to compare the mean of two sets brain structural images and identify differentiating brain regions.}

We apply the proposed NAPA test to this dataset, and also compare it with the alternative tests, all under the nominal level $\alpha = 0.05$. The number of identified differential locations by NAPA, BH, GAP, LAWS, and their simple combination {is 1193, 349, 641, 539, and 948,} respectively. Besides, the set of locations found by NAPA contains the majority of those found by BH, GAP, LAWS, and their simple combination, with the overlapping percentage equal to {97.1\% of BH, 90.8\% of GAP, 95.4\% of LAWS, and 82.4\%} of the simple combination. Again, together with our simulation studies, it seems to suggest that our proposed NAPA test manages to achieve the best power. Comparing the identified locations with the Desikan-Killiany brain atlas \citep{Desikan2006}, a number of brain regions stand out, including the left and right entorhinal cortex, the left and right posterior cingulate cortex, and left precuneus, among others. These findings generally agree with the current literature on ADHD. Particularly, the posterior cingulate cortex forms a central node in the default mode network of the brain, and has been shown to communicate with various brain networks. ADHD has been suggested as a disorder of the default mode network, and there has been evidence showing that abnormalities in the posterior cingulate cortex may disrupt the default mode network that leads to attentional lapses \citep{Nakao2011}.

\section*{Supplementary Materials}
The Supplementary Appendix contains all proofs and additional numerical results and discussions. 
\par



\bibhang=1.7pc
\bibsep=2pt
\fontsize{9}{14pt plus.8pt minus .6pt}\selectfont
\renewcommand\bibname{\large \bf References}
\expandafter\ifx\csname
natexlab\endcsname\relax\def\natexlab#1{#1}\fi
\expandafter\ifx\csname url\endcsname\relax
\def\url#1{\texttt{#1}}\fi
\expandafter\ifx\csname urlprefix\endcsname\relax\def\urlprefix{URL}\fi

\bibliographystyle{apalike}      
\bibliography{ref-napa}   

\end{document}